\documentclass[twocolumn]{aastex63}
\usepackage{graphicx}
\usepackage{aas_macros}
\usepackage{amsmath}
\usepackage{apjfonts}
\usepackage{CJKutf8}

\DeclareMathAlphabet{\mathsc}{OT1}{cmr}{m}{sc}
\def\testbx{bx}%
\DeclareRobustCommand{\ion}[2]{%
\relax\ifmmode
\ifx\testbx\f@series
{\mathbf{#1\,\mathsc{#2}}}\else
{\mathrm{#1\,\mathsc{#2}}}\fi
\else\textup{#1\,{\mdseries\textsc{#2}}}%
\fi}

\begin{document}

\title{An Imprint of the Galactic Magnetic Field in the Diffuse Unpolarized Dust Emission }

\author[0000-0001-7449-4638]{Brandon S. Hensley}
\affiliation{Spitzer Fellow, Department of Astrophysical Sciences, Princeton University, Princeton, NJ 08544, USA}
\email{bhensley@astro.princeton.edu}

\author[0000-0001-8288-5823]{Cheng Zhang \begin{CJK*}{UTF8}{gbsn}(张程)\end{CJK*}} 
\affiliation{California Institute of Technology, 1200 E California Blvd, Pasadena, CA 91125, USA}

\author{James J. Bock}
\affiliation{California Institute of Technology, 1200 E California Blvd, Pasadena, CA 91125, USA}
\affiliation{Jet Propulsion Laboratory, California Institute of Technology, Pasadena, CA 91109, USA}

\date{\today}
             
\begin{abstract}
    It is well known that aligned, aspherical dust grains emit polarized radiation and that the degree of polarization depends on the angle $\psi$ between the interstellar magnetic field and the line of sight. However, anisotropy of the dust absorption cross sections also modulates the {\it total intensity} of the radiation as the viewing geometry changes. We report a detection of this effect in the high Galactic latitude {\it Planck}  data, finding that the 353\,GHz dust intensity per $N_{\ion{H}{i}}$ is smaller when the Galactic magnetic field is mostly in the plane of the sky and larger when the field is mostly along the line of sight. These variations are of opposite sign and roughly equal magnitude as the changes in polarized intensity per $N_{\ion{H}{i}}$ with $\psi$, as predicted. In principle, the variation in intensity can be used in conjunction with the dust polarization angle to constrain the full 3D orientation of the Galactic magnetic field.
\end{abstract}

 
\section{Introduction} 
From the discovery of starlight polarization induced by dichroic extinction in the interstellar medium \citep[ISM,][]{Hiltner_1949,Hall_1949} and articulation of a theory of grain alignment \citep{Davis+Greenstein_1951, Spitzer+Tukey_1951}, measurements of dust extinction and emission have been recognized as a way to trace magnetic fields \citep{Hiltner_1951,Stein_1966}. While historically starlight polarization has been and continues to be a powerful probe of magnetic fields in the interstellar medium \citep[e.g.,][]{Clemens+etal_2012,Planck_Int_XXI,Panopoulou+etal_2019}, the relatively recent advent of sensitive ground-based, stratospheric, and space-based probes of polarized far-infrared (FIR) dust emission has opened new windows to study magnetic fields in molecular clouds, protoplanetary disks, and the diffuse ISM \citep[e.g.,][]{Benoit+etal_2004,Galitzki+etal_2014,Stephens+etal_2014,Planck_Int_XIX,WardThompson+etal_2017,Chuss+etal_2019}. In particular, maps of polarized dust emission from the {\it Planck} satellite enable mapping of magnetic fields across the entire sky \citep{Planck_Int_XX,Planck_2018_XII}.

\begin{figure*}
\centering
\includegraphics[trim=0 250 0 150,clip, width=\textwidth]{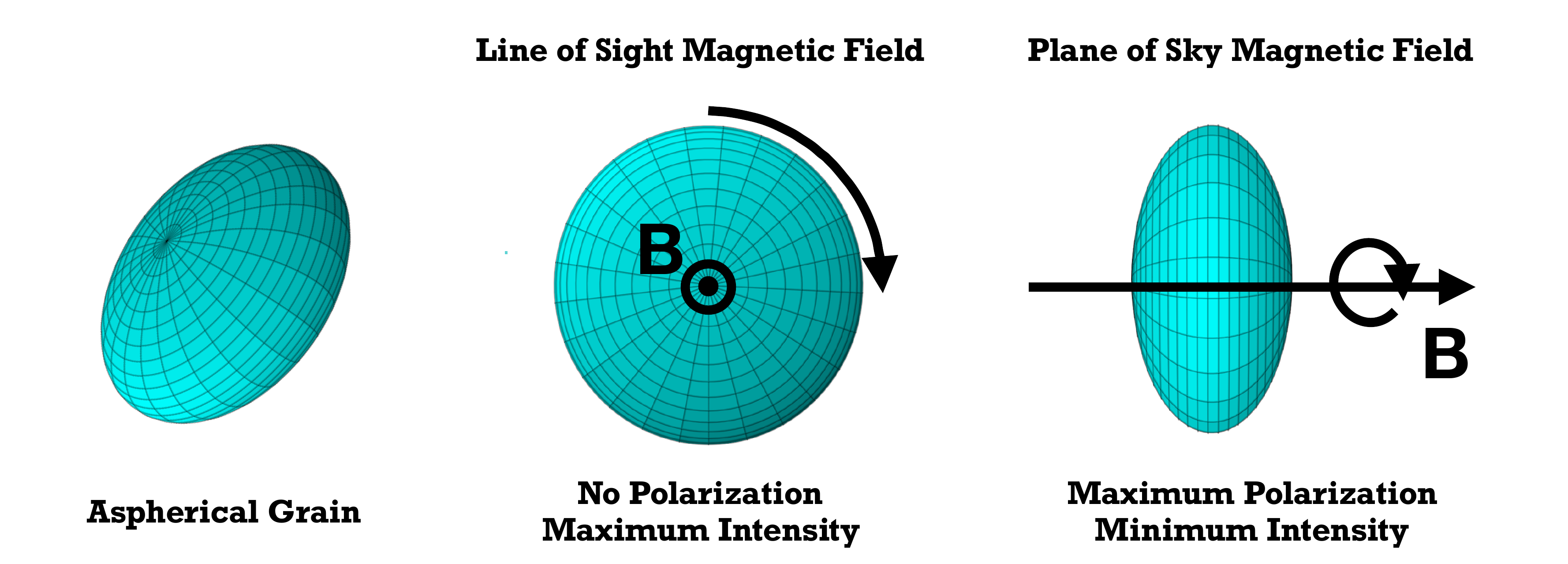}
\caption{Schematic illustrating the effect of orientation on dust intensity and polarization. Grains rotate about their shortest principal axis, and the rotation axis is systematically aligned with the interstellar magnetic field. Therefore, the grain has a larger effective cross section, and thus greater emission, when the magnetic field is oriented along the line of sight than in the plane of the sky. In contrast, grain rotation eliminates polarization when the magnetic field is along the line of sight, while polarization is maximal when the magnetic field is in the plane of the sky.}
\label{fig:schematic}
\end{figure*}

Because of the importance of the Galactic magnetic field in many areas of astrophysics, including cosmic ray propagation, the formation and structure of the Milky Way, and polarized foregrounds for Cosmic Microwave Background (CMB) science, constraining its morphology has been the subject of much research \citep[see][for a recent review]{Jaffe_2019}. With the sensitive, full-sky {\it Planck} maps, dust emission is actively being used to complement other tracers, such as synchrotron emission and pulsar rotation measures, yielding insights on both the large-scale field morphology and the Local Bubble \citep[e.g.,][]{Planck_Int_XLII,Alves+etal_2018,Pelgrims+etal_2018}.

However, the dust polarization angle as determined from measurements of the Stokes $Q$ and $U$ parameters is sensitive only to the plane of sky component of the Galactic magnetic field. In this work, we describe a new method for constraining the line of sight component of the magnetic field, and thus the full 3D orientation, using the total (i.e., unpolarized) intensity and the \ion{H}{i} column density.

Emission from interstellar dust grains dominates the spectral energy distribution (SED) of the Galaxy from FIR to microwave wavelengths. The interstellar magnetic field aligns dust grains such that their short axes are parallel to the field direction \citep[see][for a review]{Andersson+etal_2015}. Since grains emit photons preferentially with electric field oriented along the long axis of the grain, the resulting emission from a population of dust grains is polarized. Observations of polarized dust emission allow straightforward determination of the orientation of the interstellar magnetic field projected onto the plane of the sky. When the magnetic field lies in the plane of the sky, the polarization is maximal, while the polarization vanishes when the magnetic field is parallel to the line of sight. This effect accounts for much of the variation in the dust polarization fraction across the sky, though averaging of multiple magnetic field orientations along the line of sight is also an important contributor \citep{Clark_2018}.

While the effect of the viewing angle is most readily apparent in polarization, in principle the total intensity of dust emission also depends on the angle between the line of sight and the interstellar magnetic field \citep{Lee+Draine_1985}. This arises simply because the effective cross section of an aspherical grain changes with viewing angle. Consequently, the effective cross section of dust grains, and thus the total emission per grain, is larger when the magnetic field is along the line of sight, and smaller when it lies in the plane of the sky, as illustrated in Figure~\ref{fig:schematic}. Using the {\it Planck} 353\,GHz data, we present the first detection of this effect in intensity.

In addition to mapping magnetic fields, the dust polarization angles and their statistics provide tests of models of magnetohydronamic (MHD) turbulence. For instance \citet{Planck_2018_XII} found that the dispersion in polarization angles is best reproduced by models with strong turbulence. In this work, we demonstrate that the polarization angle dispersion function $\mathcal{S}$ is strongly influenced by the orientation of the Galactic magnetic field with respect to the line of sight, which may complicate this interpretation.

This paper is organized as follows: in Section~\ref{sec:theory}, we review the physics of emission from partially aligned dust grains and derive our central relationship between the total and polarized intensity per $N_\ion{H}{i}$; in Section~\ref{sec:data}, we describe the various data products used in this study; we discuss the statistical model employed in our analysis in Section~\ref{sec:model}; we present a detection of the correlation between the dust emission and the magnetic field geometry in Section~\ref{sec:results}; in Section~\ref{sec:discussion}, we discuss the significance of this result for both mapping the Galactic magnetic field and for modeling polarized dust foregrounds; and we summarize our conclusions in Section~\ref{sec:conclusions}.

\section{Theory}
\label{sec:theory}

\subsection{Emission and Polarization from Aligned Grains}
\label{subsec:dust_emission}
While the actual shapes of interstellar grains are unknown, they are often modeled as spheroids or ellipsoids. For generality, we model dust grains as triaxial ellipsoids with principal axes $a_1 \leq a_2 \leq a_3$. Let $\hat{\bf a}_1$, $\hat{\bf a}_2$, and $\hat{\bf a}_3$ be unit vectors along the respective axes. Suprathermally rotating interstellar grains spin about their short axis, i.e., angular momentum ${\bf J} \parallel \hat{\bf a}_1$. Paramagnetic dissipation brings ${\bf J}$ into alignment with the interstellar magnetic field ${\bf B}$, and so perfectly aligned grains have $\hat{\bf a}_1 \parallel {\bf B}$. 

In this work, we model polarized dust emission at $\lambda = 850$\,$\mu$m, which is much larger than the size of interstellar grains ($a \lesssim 0.1$\,$\mu$m). In the limit of $a/\lambda \ll 1$, i.e., the ``electric dipole limit,'' the polarization cross section of an arbitrarily oriented grain can be written in terms of the cross sections for photons propagating along the principal axes. For non-magnetic grains in the electric dipole limit, this reduces to just the three cross sections $C_{\rm abs}^1$, $C_{\rm abs}^2$, and $C_{\rm abs}^3$, which are the absorption cross sections for photons polarized with ${\bf E}$ parallel to $\hat{\bf a}_1$, $\hat{\bf a}_2$, and $\hat{\bf a}_3$, respectively. In the limit of perfect internal alignment, grain rotation further simplifies the analysis by time averaging $C_{\rm abs}^2$ and $C_{\rm abs}^3$ irrespective of the grain orientation.

Under these assumptions, the absorption cross section for randomly oriented grains $C_{\rm abs}^{\rm ran}$ is given by

\begin{equation}
    C_{\rm abs}^{\rm ran} = \frac{1}{3}\left(C_{\rm abs}^1 + C_{\rm abs}^2 + C_{\rm abs}^3\right)
    ~~~.
\end{equation}
A population of randomly-oriented grains produces no net polarization.

Now consider a population of grains with $\hat{\bf a}_1$ perfectly aligned with ${\bf B}$. Let $\psi$ be the angle between the line of sight and ${\bf B}$. The total and polarized absorption cross sections can be defined relative to two orthogonal polarization modes:

\begin{align}
    C_{\rm abs} &\equiv \frac{1}{2}\left(C_{\rm abs}^\perp + C_{\rm abs}^\parallel\right) \\
    C_{\rm pol} &\equiv \frac{1}{2}\left(C_{\rm abs}^\perp - C_{\rm abs}^\parallel\right)
    ~~~.
\end{align}
We define $C_{\rm abs}^\perp$ and $C_{\rm abs}^\parallel$ relative to the orientation of ${\bf B}$ projected onto the plane of the sky. With this definition, perfectly aligned grains have total and polarized absorption cross sections

\begin{align}
    C_{\rm abs}^{\rm align} &= \frac{1}{2}\left(C_{\rm abs}^2 + C_{\rm abs}^3\right)\cos^2\psi + \frac{1}{2}\left[C_{\rm abs}^1 + \frac{1}{2}\left(C_{\rm abs}^2 + C_{\rm abs}^3\right)\right]\sin^2\psi \\
    C_{\rm pol}^{\rm align}  &= \frac{1}{2}\left[\frac{1}{2}\left(C_{\rm abs}^2 + C_{\rm abs}^3\right) - C_{\rm abs}^1\right]\sin^2\psi
    ~~~,
\end{align}
respectively.

Grain alignment and disalignment mechanisms operate on finite timescales and so in any environment there will be a distribution of angles between the grain angular momenta ${\bf J}$ and ${\bf B}$, with ${\bf J}$ tending to precess around ${\bf B}$. However, in the electric dipole limit, this can be accurately approximated by assuming a fraction $f$ of grains are perfectly aligned with ${\bf B}$ and the remaining $(1-f)$ are randomly oriented \citep{Dyck+Beichman_1974}. Thus,

\begin{align}
    C_{\rm abs} &= f C_{\rm abs}^{\rm align} + \left(1-f\right)C_{\rm abs}^{\rm ran} \\
    C_{\rm pol} &= f C_{\rm pol}^{\rm align}
    ~~~.
\end{align}

In the electric dipole limit, the absorption cross section per unit grain volume $V$ is independent of grain size. Assuming the grains have mass density $\rho$, it is convenient to define the opacities

\begin{align}
    \kappa_\nu^{\rm ran} &= \frac{1}{3\rho V}\left(C_{\rm abs}^1 + C_{\rm abs}^2 + C_{\rm abs}^3\right) \\
    \kappa_\nu^{\rm pol} &= \frac{1}{2\rho V}\left[\frac{1}{2}\left(C_{\rm abs}^2 + C_{\rm abs}^3\right) - C_{\rm abs}^1\right]
\end{align}
such that a population of grains with mass surface density $\Sigma_d$ and temperature $T_d$ emits total and polarized intensities

\begin{align}
    \label{eq:INH}
    I_\nu &= \Sigma_d B_\nu\left(T_d\right) \left[\kappa_\nu^{\rm ran} + f\kappa_\nu^{\rm pol}\left(\frac{2}{3} - \sin^2\psi\right)\right] \\
    \label{eq:PNH}
    P_\nu &= \Sigma_d B_\nu\left(T_d\right) f \kappa_\nu^{\rm pol}\sin^2\psi
~~~,
\end{align}
where $B_\nu\left(T\right)$ is the Planck function. The dust polarization fraction $p_\nu$ is defined as

\begin{equation}
\label{eq:pfrac}
    p_\nu \equiv \frac{P_\nu}{I_\nu} = \frac{f\sin^2\psi}{1 + \frac{\kappa_\nu^{\rm pol}}{\kappa_\nu^{\rm ran}}f\left(\frac{2}{3} - \sin^2\psi\right)}\frac{\kappa_\nu^{\rm pol}}{\kappa_\nu^{\rm ran}}
    ~~~.
\end{equation}

Since we employ $N_\ion{H}{i}$ as a proxy for the dust column, we define the dust to gas mass ratio $\delta_{\rm DG}$ as

\begin{equation}
    \delta_{\rm DG} \equiv \frac{\Sigma_d}{m_p N_\ion{H}{i}}
    ~~~,
\end{equation}
where $m_p$ is the proton mass. Combining these equations, we obtain our principal result:

\begin{equation}\label{eq:main}
    \frac{I_\nu}{N_\ion{H}{i}} = \left(\kappa_\nu^{\rm ran} + \frac{2f}{3}\kappa_\nu^{\rm pol}\right)m_p \delta_{\rm DG} B_\nu\left(T_d\right) - \frac{P_\nu}{N_\ion{H}{i}}
    ~~~.
\end{equation}
If the dust properties $\kappa_\nu^{\rm ran}$, $\kappa_\nu^{\rm pol}$, $f$ $\delta_{\rm DG}$, and $T_d$ are not varying across the sky, we expect a negative correlation between the observed $I_\nu/N_\ion{H}{i}$ and $P_\nu/N_\ion{H}{i}$ with slope of -1.

Instead of working with the limiting cases of perfect alignment and random orientation, the effects of imperfect grain alignment can be parameterized by the Rayleigh reduction factor $\mathcal{R}$ \citep{Greenberg_1968}. $\mathcal{R}$ accounts for the distribution of angles $\theta$ between ${\bf J}$ and ${\bf B}$ and is given by

\begin{equation}
    \mathcal{R} \equiv \frac{3}{2}\left(\langle\cos^2\theta\rangle - \frac{1}{3}\right)
    ~~~.
\end{equation}
In the case of both oblate ($C_{\rm abs}^2 = C_{\rm abs}^3$) and prolate ($C_{\rm abs}^1 = C_{\rm abs}^2$) spheroidal grains, the equations for the total and polarized intensities in the electric dipole limit are recovered from our results by simply replacing the alignment fraction $f$ with $\mathcal{R}$ \citep{Lee+Draine_1985,Draine+Hensley_2016}.

This derivation assumes all dust emission is associated with a single magnetic field orientation. In reality, averaging of multiple orientations both along the line of sight and within the beam is expected. These effects typically reduce the level of polarization and bring $I_\nu/N_\ion{H}{i}$ closer to the mean value.

\subsection{The Polarization Angle Dispersion Function}
\label{subsec:theory_S}

The polarized emission from interstellar dust is characterized by the Stokes parameters $Q_\nu$ and $U_\nu$, which are directly observable. When expressed as specific intensities, they are related to the polarized intensity by

\begin{align}
    Q_\nu &= P_\nu\cos 2\chi \\
    U_\nu &= P_\nu\sin 2\chi
    ~~~,
\end{align}
where $\chi$ is the polarization angle and Stokes $V$ has been assumed to be zero. For electric dipole emission from dust grains aligned with $\hat{\bf a}_1 \parallel {\bf B}$, $\chi$ is the angle perpendicular to the projection of ${\bf B}$ onto the plane of the sky. The numerical value of $\chi$ depends on the adopted polarization convention.

The polarization angle dispersion function $\mathcal{S}$ is a measure of the spatial variability of $\chi$ over a given region. For a given pixel $i$ on the sky, $\mathcal{S}$ is computed over a region $R$ such that

\begin{equation}
    \label{eq:S}
    \mathcal{S}_i = \sqrt{\frac{1}{N}\sum_{j\in R} \left(\chi_j - \chi_i\right)^2}
    ~~~,
\end{equation}
where the sum is over the $N$ pixels in $R$. In practice, $R$ is often taken to be annulus centered on pixel $i$ of width equal to the radius of its inner edge (see Section~\ref{subsec:data_S}). Note that $\mathcal{S}$ is {\it not} dependent on the adopted polarization convention, though care should be taken to avoid computing $\mathcal{S}$ near singularities in the coordinate system \citep[see discussion in][]{Planck_Int_XIX}.

Interpretation of $\mathcal{S}$ can be subtle as a number of effects can give rise to dispersion in the polarization angle. For instance, $\mathcal{S}$ will be high in a region in which the magnetic field is relatively disordered. Indeed, measurements of $\mathcal{S}$ (and the closely related structure function) have been used to constrain models of interstellar turbulence in clouds and in the diffuse ISM \citep[e.g.][]{Hildebrand+etal_2009,Houde+etal_2009,Poidevin+etal_2013,Planck_Int_XX,Planck_2018_XII}. Disorder in the magnetic field along the line of sight can likewise induce spatial variability in the measured polarization angles, as can instrumental noise in the $Q_\nu$ and $U_\nu$ measurements.

The magnitude of each of these effects is modulated by the angle $\psi$ between the line of sight and ${\bf B}$ \citep{FalcetaGoncalves+etal_2008,Poidevin+etal_2013,King+etal_2018}. If ${\bf B}$ is nearly along the line of sight, then even small perturbations to the 3D field orientation can induce large changes in $\chi$, leading to large values of $\mathcal{S}$. In contrast, if ${\bf B}$ lies mostly in the plane of the sky, small perturbations to its 3D orientation have only a modest impact on $\chi$, resulting in small values of $\mathcal{S}$. This connection between $\mathcal{S}$ and $\psi$ is consistent with the strong empirical anticorrelation between $\mathcal{S}$ and the dust polarization fraction $p_\nu$ observed over a large range of column densities \citep{Planck_Int_XIX,Fissel+etal_2016,Planck_2018_XII}.

The $p_\nu$--$\mathcal{S}$ relation is often parameterized with a power law. Over the full sky, \citet{Planck_2018_XII} found that $p_\nu \propto \mathcal{S}^{-1}$ while \citet{Fissel+etal_2016} found that $p_\nu \propto \mathcal{S}^{-0.67}$ in the Vela~C Molecular Cloud. Given these results, we hypothesize that over the diffuse high latitude sky, 

\begin{equation}
    \sin^2\psi \propto \mathcal{S}^n
    ~~~.
\end{equation}
Under this assumption, Equations~\ref{eq:INH} and \ref{eq:PNH} can be rewritten as

\begin{align}
    \label{eq:I_S}
    \frac{I_\nu}{N_\ion{H}{i}} &= A - B \left(\frac{\mathcal{S}}{1^\circ}\right)^n \\
    \label{eq:P_S}
    \frac{P_\nu}{N_\ion{H}{i}} &= B \left(\frac{\mathcal{S}}{1^\circ}\right)^n
    ~~~,
\end{align}
where $A$, $B$, and $n$ are global mean values with scatter induced by variations in dust properties, such as the dust temperature and dust to gas ratio. In this work, we find that the observed relationship between $I_\nu$ and $P_\nu$ in the low column density sky ($N_\ion{H}{i} < 4\times10^{20}\,$cm$^{-2}$) is well described by this model.

\section{Data}
\label{sec:data}

In this section we introduce the data sets used in this work. Our principal analysis is performed on HEALPix\footnote{\url{https://healpix.sourceforge.io/}} \citep{Gorski+etal_2005} maps smoothed to a resolution of $160'$ with $N_{\rm side} = 64$.

\subsection{Dust Emission Maps}
The 353\,GHz band was the highest-frequency {\it Planck} channel designed for polarimetry. Dominated by thermal dust emission, it is ideal for studying dust polarization properties across the sky.

Like \citet{Planck_2018_XII}, we are interested in the astrophysical properties of Galactic dust and therefore require maps minimally contaminated by other sources of emission, notably the CMB and the Cosmic Infrared Background (CIB). We therefore follow \citet{Planck_2018_XII} in utilizing the component-separated maps produced by application of the Generalized Needlet Internal Linear Combination (GNILC) algorithm \citep{Remazeilles+etal_2011} to the {\it Planck} data \citep{Planck_2018_IV}. Critically for this study, these maps were created with the intention of separating the contributions of the CIB and Galactic dust by exploiting both spatial and spectral information. This is not true of the dust maps made with the {\texttt Commander} parametric component separation algorithm, in which both CIB and Galactic dust emission are absorbed in the ``thermal dust'' component \citep{Planck_2015_X,Planck_2018_IV}.

We employ the 2018 GNILC Stokes $I$, $Q$, and $U$ maps\footnote{COM\_CompMap\_IQU-thermaldust-gnilc-unires\_2048\_R3.00.fits} at 353\,GHz. We smooth the maps to $160'$ resolution and repixellate them at $N_{\rm side} = 64$. Following \citet{Planck_2018_XII}, we subtract 389\,$\mu$K$_{\rm CMB}$ from the $I$ map to account for zero level offsets from, e.g, the CIB monopole, and convert to MJy\,sr$^{-1}$ with the factor 287.5\,MJy\,sr$^{-1}$\,$\mu$K$_{\rm CMB}^{-1}$ \citep{Planck_2018_III}. The $P$ map is obtained via $P = \sqrt{Q^2 + U^2}$. Because of the substantial smoothing, the noise bias is very small even at low column densities \citep{Planck_2018_XII}, and so we do not perform any debiasing.

\subsection{Polarization Angle Dispersion Function}
\label{subsec:data_S}
Employing the 353\,GHz GNILC $Q$ and $U$ maps, \citet{Planck_2018_XII} constructed maps of the polarization angle dispersion function $\mathcal{S}$ (see Equation~\ref{eq:S}) at different resolutions. In all cases, they computed $\mathcal{S}$ on an annulus with inner radius $\delta/2$ and outer radius $3\delta/2$, where the ``lag'' $\delta$ is taken to be half the resolution of the map. They found that the map of $\mathcal{S}$ computed at $80'$ resolution ($\delta = 40'$) was affected by noise bias and so recommended resolution $160'$ ($\delta = 80'$) when computing $\mathcal{S}$. In this work, we employ their map\footnote{V. Guillet, private comm.} of $\mathcal{S}$ at $160'$ pixellated at $N_{\rm side}=64$. With this resolution, we find that debiasing would have a 10\% or greater effect on the value of $\mathcal{S}$ in less than 1\% of the high latitude pixels considered in this work. Therefore, as with the $P$ map, we do not perform any debiasing, and we have verified that this choice does not affect any of our conclusions. The $\mathcal{S}$ map requires the coarsest resolution of all data products analyzed, and so we smooth all other maps to this resolution.

\subsection{\ion{H}{i} Map}
\label{sec:dataNHI}

The 21\,cm line from atomic H has been mapped spectroscopically over the Southern Hemipshere by the Parkes Galactic All-Sky Survey \citep[GASS,][]{McClureGriffiths+etal_2009} and the Northern Hemisphere by the Effelsberg-Bonn $\ion{H}{i}$ Survey \citep[EBHIS,][]{Kerp+etal_2011}. These datasets have been combined and homogenized by the HI4PI Survey \citep{HI4PI} to a uniform $16\overset{'}{.}2$ resolution over the full sky. In this work, we employ the $N_\ion{H}{i}$ map\footnote{\url{https://doi.org/10.7910/DVN/AFJNWJ}} derived from the HI4PI data by \citet{Lenz+etal_2017}, which filters out all \ion{H}{i} emission having radial velocity $|v_{\rm LSR}| > 90$\,km\,s$^{-1}$. Such high velocity gas was found to have little correlation with dust reddening \citep{Lenz+etal_2017}, and so the filtered map provides a better predictor of the dust column. Additionally, this velocity-filtered map is free from extragalactic contamination down to very low levels \citep{Chiang+Menard_2019}. We smooth the \ion{H}{i} map to a resolution of $160'$ and repixellate to $N_{\rm side} = 64$.

All analysis in this work is restricted to lines of sight with $N_\ion{H}{i} < 4\times10^{20}$\,cm$^{-2}$, where the \ion{H}{i} column density has been shown to be a linearly related to the dust column density \citep{Lenz+etal_2017}. At $160'$ resolution and $N_{\rm side} = 64$, this corresponds to a sky fraction of 39\% and 19,120 pixels.

\subsection{Dust Temperature Map}
\label{subsec:data_Td}
In addition to its application to the {\it Planck} data, the GNILC algorithm has also been employed on the IRIS 100\,$\mu$m map \citep{MivilleDeschenes+Lagache_2005}, a reprocessing of the IRAS 100\,$\mu$m map \citep{Wheelock+etal_1994}, to derive multi-frequency Galactic dust SEDs across the sky \citep{Planck_Int_XLVIII}. The {\it Planck} and IRIS Galactic dust maps were then fit with a modified blackbody emission model to derive full-sky maps of dust temperature $T_d$ that are minimally contaminated by CIB emission. We employ the resulting $T_d$ map\footnote{COM\_CompMap\_Dust-GNILC-Model-Temperature\_2048\_R2.00.fits}, which has a native resolution of approximately $5'$ and is pixellated with $N_{\rm side} = 2048$. We smooth this map to $160'$ resolution and repixellate to $N_{\rm side} = 64$. 

In principle, maps of $T_d$ should not be smoothed by simple averaging but rather by redoing the parameter fits on smoothed intensity maps. We investigate the potential impact of smoothing on $T_d$ in Section~\ref{subsec:Tdust} and find our principal conclusions relating to $T_d$ also hold at $16\overset{'}{.}2$ resolution, suggesting that such effects are not important for our analysis.

\section{Data Model}
\label{sec:model}

\subsection{Statistical Model}
\label{subsec:stat_model}
The principal aim of this work is to assess whether the observed 353\,GHz dust intensity per $N_\ion{H}{i}$ has a statistically significant correlation with the orientation of the Galactic magnetic field relative to the line of sight. The maps of the dust intensity and \ion{H}{i} intensity employed in this study have high signal-to-noise, and the maps of the polarized intensity and polarization angle dispersion $\mathcal{S}$ are also signal dominated because of the aggressive smoothing. However, $I_\nu/N_\ion{H}{i}$ and $P_\nu/N_\ion{H}{i}$ are expected to have large astrophysical scatter owing to changes in, e.g, the dust to gas ratio, dust composition, and dust temperature in addition to the orientation effect we seek.

In Section~\ref{subsec:I_P}, we investigate the relationship between $I_\nu/N_\ion{H}{i}$ and $P_\nu/N_\ion{H}{i}$ predicted by Equation~\ref{eq:main}. While the magnetic field orientation induces a negative correlation with slope $m = -1$, variations in other properties such as dust temperature and dust to gas ratio induce a positive correlation. To test whether a statistically significant negative correlation is present in the data, we employ the likelihood function

\begin{equation}
\label{eq:L_IP}
    \mathcal{L} \propto \prod_i \frac{1}{\sigma_I}\,{\rm exp}\left[-\frac{\left(I_\nu/N_{\ion{H}{i},i} - m P/N_{\ion{H}{i},i} - b\right)^2}{\sigma_I^2}\right]
    ~~~,
\end{equation}
where we have assumed a linear relationship with slope $m$ and intercept $b$. $\sigma_I$ is the random scatter in the $I_\nu/N_\ion{H}{i}$ measurements and is assumed to be constant across the sky. While this model enables determination of whether or not a trend exists between $I_\nu/N_\ion{H}{i}$ and $P_\nu/N_\ion{H}{i}$, it does not take into account that pixel-to-pixel variations in these two quantities can be strongly correlated. We discuss these correlated variations in Section~\ref{subsec:I_P}.

In Section~\ref{subsec:I_S}, we employ Equations~\ref{eq:I_S} and \ref{eq:P_S} to model the relations between $I_\nu/N_\ion{H}{i}$, $P_\nu/N_\ion{H}{i}$, and $\mathcal{S}$. We fit for the parameters $A$, $B$, and $n$ in the model using the observations of both $I_\nu/N_\ion{H}{i}$ and $P_\nu/N_\ion{H}{i}$. We employ the likelihood function

\begin{align}
    \label{eq:L_IPS}
    \mathcal{L} \propto \prod_i \frac{1}{\sigma_I\sigma_P}\,&{\rm exp}\left[-\frac{\left(I_{\nu,i}/N_{\ion{H}{i},i} - A + B\mathcal{S}_i^n\right)^2}{\sigma_I^2}\right] \times \nonumber \\ 
    &{\rm exp}\left[- \frac{\left(P_{\nu,i}/N_{\ion{H}{i},i} - B\mathcal{S}_i^n\right)^2}{\sigma_P^2}\right]
    ~~~,
\end{align}
where $\sigma_I$ and $\sigma_P$ are the random scatter in the $I_\nu/N_\ion{H}{i}$ and $P_\nu/N_\ion{H}{i}$ measurements, respectively.

All model fitting is performed with the \texttt{emcee} Markov Chain Monte Carlo software\footnote{\url{https://emcee.readthedocs.io/en/v2.2.1/}} \citep{ForemanMackey+etal_2013}. For all parameters, we employ broad, uninformative priors.

\subsection{Dust Temperature Correction}
\label{subsec:Tdust}

\begin{figure}
\centering
\includegraphics[width=0.48\textwidth]{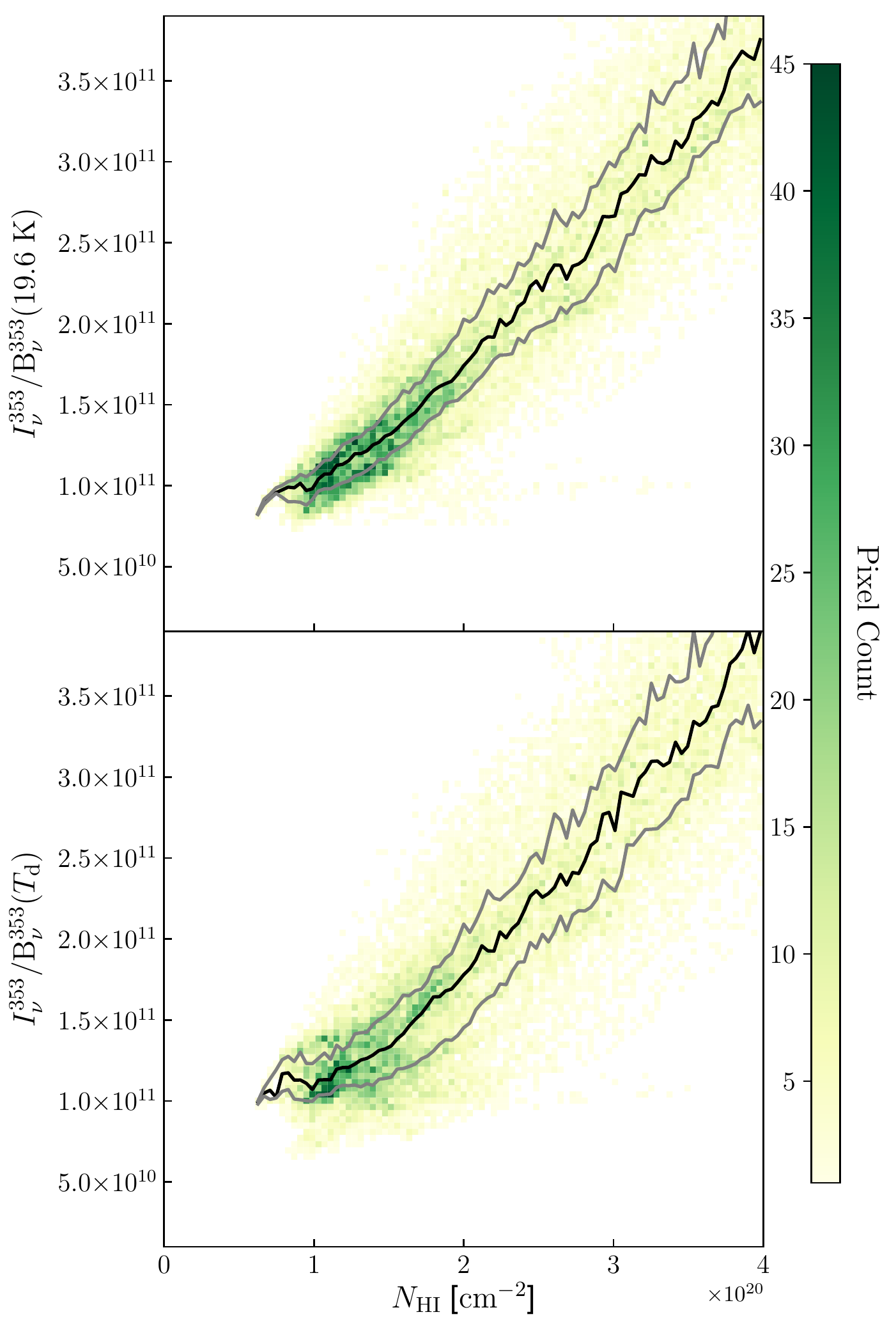}
\caption{Relationship between $N_\ion{H}{i}$ and the 353\,GHz dust intensity $I_\nu$. Top: On the y-axis, $I_\nu$ is divided by the blackbody function at $T_d = 19.6$\,K. The black line is the running median while the gray lines are the 25th and 75th percentiles. Bottom: Instead of assuming a constant $T_d$, $I_\nu$ is by the blackbody function evaluated at the $T_d$ determined by SED fits to the GNILC component separated maps \citep{Planck_Int_XLVIII}. The dust temperature correction does {\it not} improve the scatter or the linearity of the relation.}
\label{fig:in}
\end{figure}

The total and polarized emission from a fixed amount of dust are sensitive to the dust temperature (see Equations~\ref{eq:INH} and \ref{eq:PNH}), with the long-wavelength emission scaling as the first power of $T_d$ in the Rayleigh-Jeans limit. Dust temperature variations add scatter to $I_\nu/N_\ion{H}{i}$ and $P_\nu/N_\ion{H}{i}$ unrelated to orientation effects. Even more critically, $T_d$ fluctuations affect $I_\nu/N_\ion{H}{i}$ and $P_\nu/N_\ion{H}{i}$ in the same manner, inducing a {\it positive} correlation between these quantities. This effect minimizes and potentially eliminates the {\it negative} correlation expected between these quantities from orientation effects (see Equation~\ref{eq:main}). We therefore seek to mitigate the effects of dust temperature by employing a dust temperature map obtained from SED fitting. However, we find that, over this range of column densities, the available dust temperature maps do not improve upon simply assuming a constant temperature.

The correlation between the dust intensity and $N_\ion{H}{i}$ is remarkably tight, and should tighten further when a correction is made for $T_d$ variations. In Figure~\ref{fig:in}, we use the GNILC $T_d$ map to plot $I_\nu/B_\nu\left(T_d\right)$ against $N_\ion{H}{i}$, finding that the $T_d$-corrected intensity map is indeed tightly correlated with $N_\ion{H}{i}$. However, in Figure~\ref{fig:in} we present the same plot but instead assuming a sky-constant $T_d = 19.6$\,K. We find that this relationship has even less scatter, i.e., the GNILC $T_d$ map appears to make the $I_\nu$--$N_\ion{H}{i}$ correlation {\it worse}.

\begin{figure*}
\centering
\includegraphics[width=\textwidth]{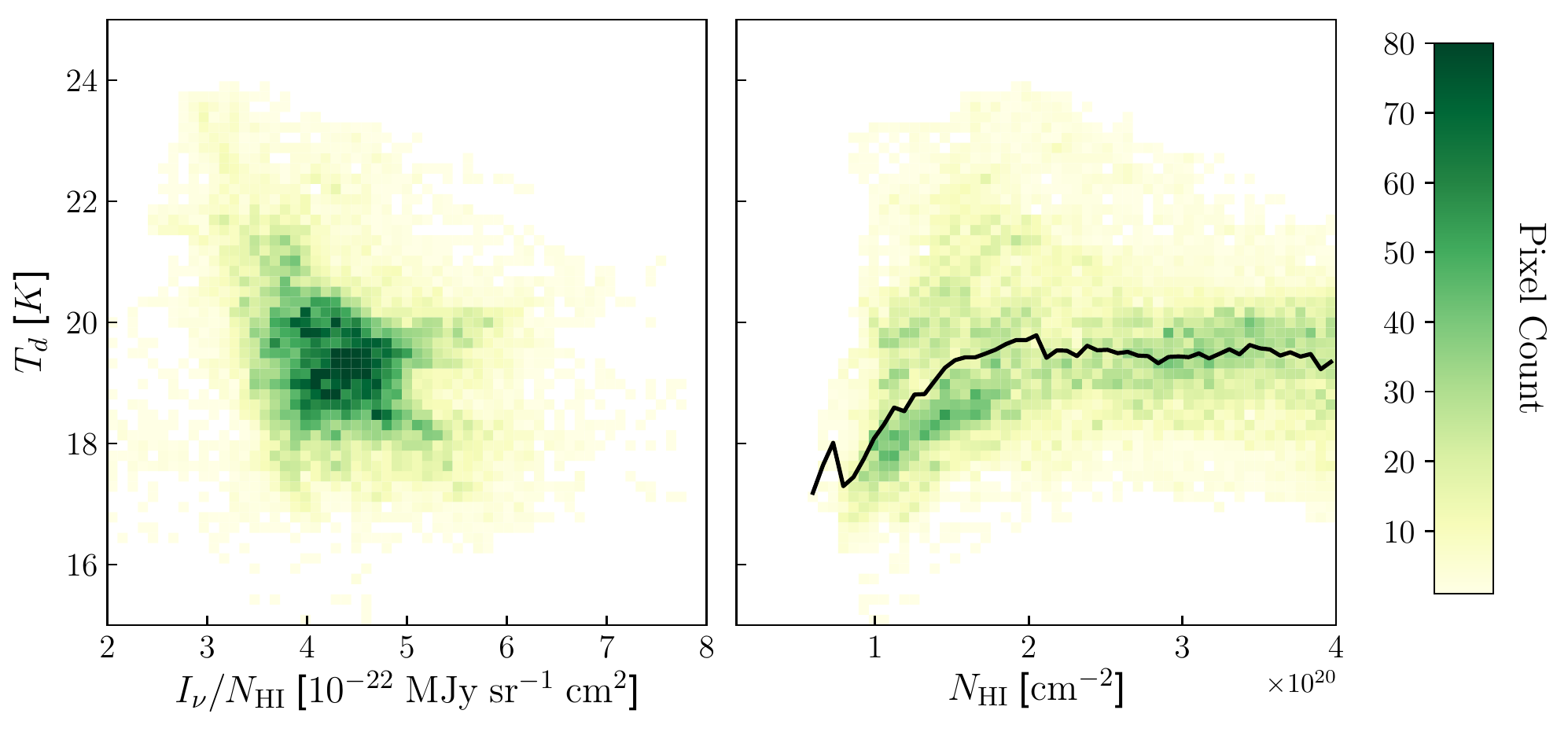}
\caption{Left: 2D histogram of $I_\nu/N_\ion{H}{i}$ versus dust temperature $T_d$. In principle, higher temperature dust should emit more radiation per H, but the expected positive correlation is not evident. Right: The relationship between the $\ion{H}{i}$ column density and dust temperature. No {\it a priori} correlation is expected in the diffuse gas examined here. However, a positive correlation is evident for $N_\ion{H}{i}$ < $2.0 \times 10^{20}\,{\rm cm^{-2}}$ as highlighted by the running median (black solid line). This may reflect a systematic trend that the lowest column density gas happens to have cooler dust temperatures, or may be an artifact of model fitting, as discussed in the text.}
\label{fig:TN}
\end{figure*}

We are examining here the lowest column densities on the sky, and therefore the regions of lowest signal-to-noise on the dust emission. It is therefore possible that variations in the $T_d$ map are driven by noise. To investigate this further, we present the correlation between $I_\nu/N_\ion{H}{i}$ and $T_d$ in the left panel of Figure~\ref{fig:TN}. We expect a strong positive correlation in this space since hotter dust produces more long-wavelength emission per grain, yet there is no apparent trend in the data. 

In contrast, as shown in the right panel of Figure~\ref{fig:TN}, we find a strong positive correlation between $N_\ion{H}{i}$ and $T_d$ for $N_\ion{H}{i} < 2\times10^{20}$\,cm$^{-2}$. While it is possible that the lowest column density sightlines have systematically cooler dust, it seems more likely that this correlation is driven by the fitting degeneracy between the dust column and the dust temperature. If the effects of $T_d$ on the shape of the dust SED cannot be adequately constrained due to low signal-to-noise, then both $T_d$ and the normalization parameter both have the effect of increasing the overall amplitude of the dust emission. Thus, an artificial correlation between $T_d$ and the true dust column can emerge. This may also explain why the effect is limited to only the very lowest values of $N_\ion{H}{i}$, where the signal-to-noise on the dust SED is the lowest.

This analysis employed the GNILC dust temperature maps smoothed to $160'$ as described in Section~\ref{subsec:data_Td}. Such aggressive smoothing, particularly by simple averaging of $T_d$ across pixels, could induce biases in the map. However, the effects seen here persist even at the limiting $16\overset{'}{.}2$ resolution of the $N_\ion{H}{i}$ map. It seems therefore unlikely that the issues identified in the $T_d$ map are solely the result of smoothing.

Finally, we attempted a dust temperature correction using the \texttt{Commander} $T_d$ map instead \citep{Planck_2015_X}. However, it is also unable to improve the $I_\nu$--$N_\ion{H}{i}$ correlation relative to assuming a constant $T_d$. In addition to being subject to similar issues as the GNILC map, this alternative dust temperature map is more prone to CIB contamination since CIB anisotropies were not fit as a separate component from Galactic dust. For a more in-depth exploration of these issues in presently-available dust temperature maps, see \citet{Herman_2019}.

We therefore conduct our analysis without a dust temperature correction, effectively assuming that $T_d$ is constant over the diffuse high-latitude sky.

\section{Results}
\label{sec:results}

\subsection{$I_\nu/N_\ion{H}{i}$--$P_\nu/N_\ion{H}{i}$ Correlation}
\label{subsec:I_P}

\begin{figure*}
\centering
\includegraphics[width=0.51\textwidth]{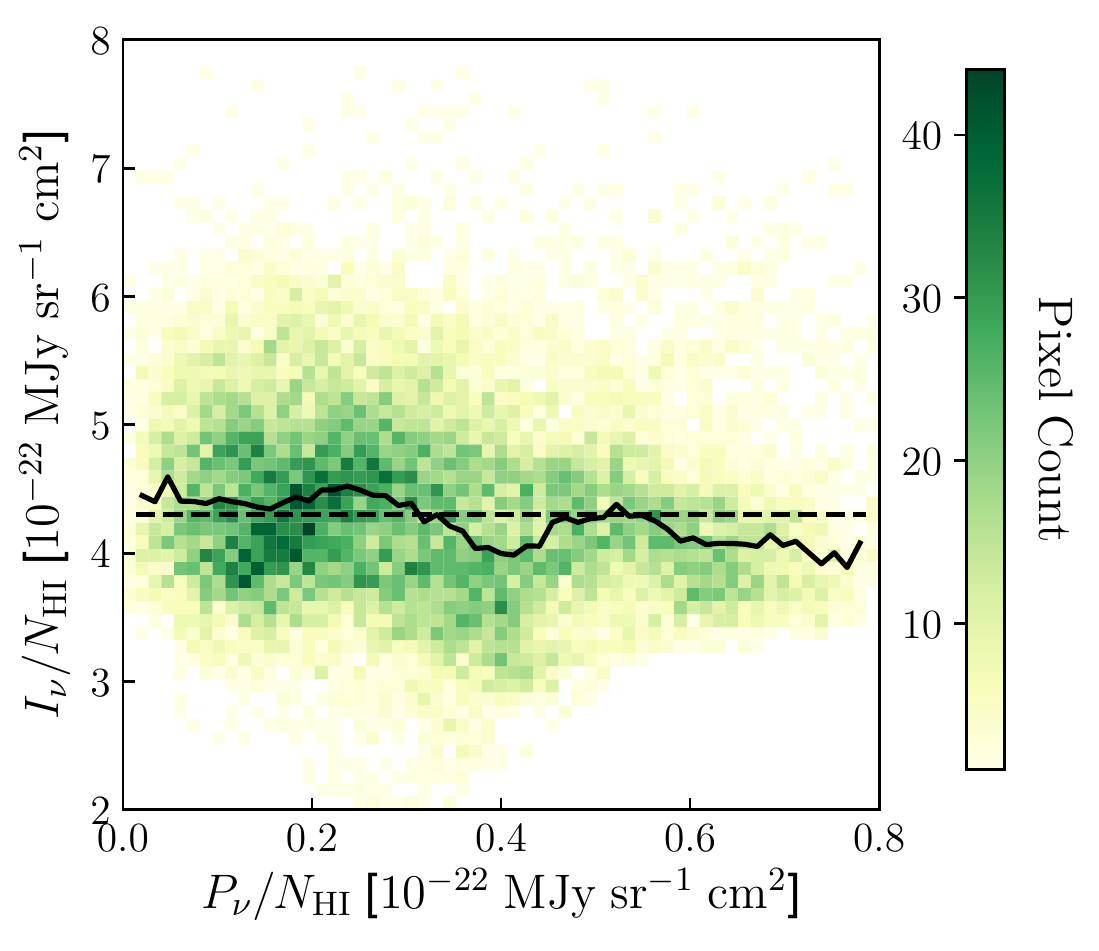}
\includegraphics[width=0.44\textwidth]{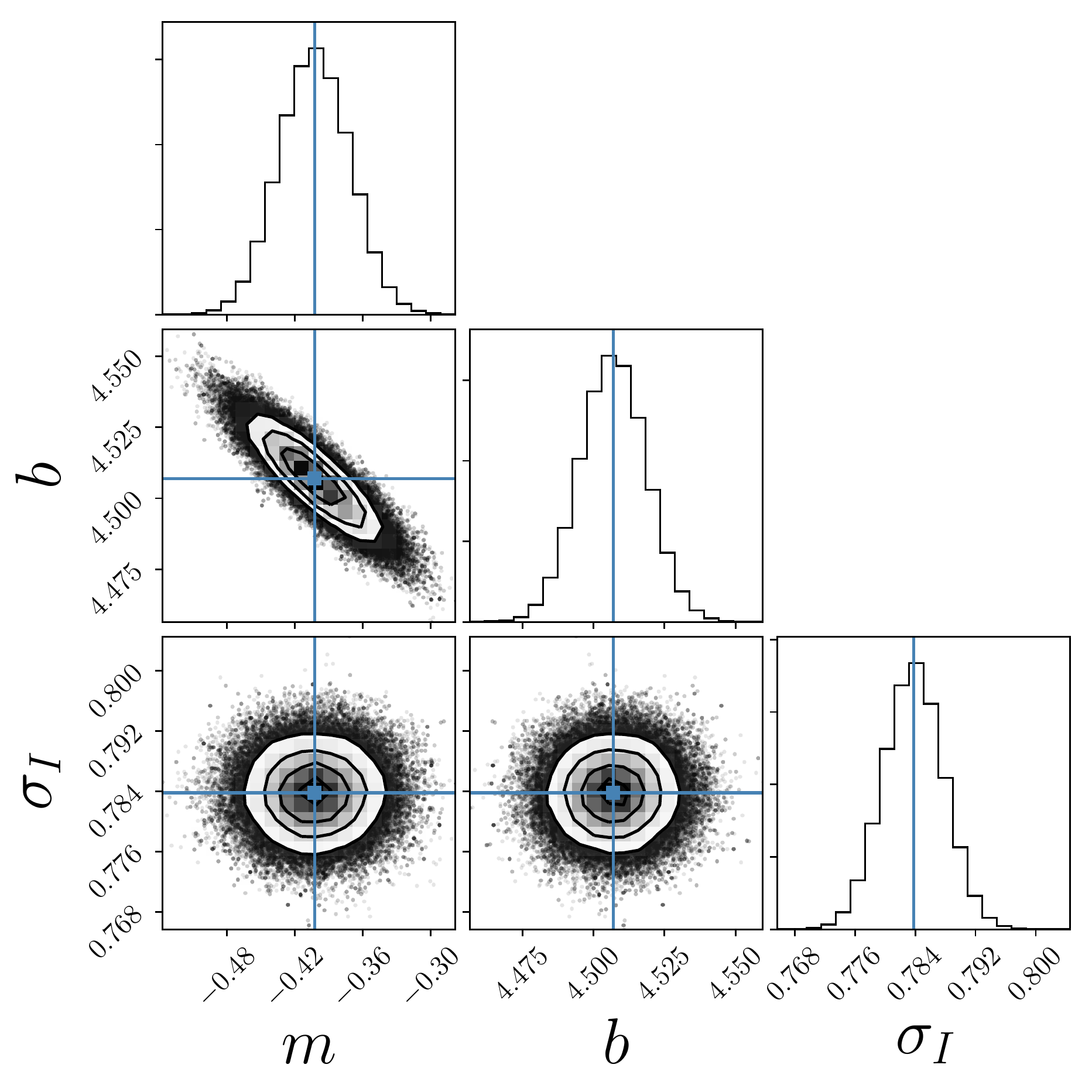}
\caption{Left: 2D histogram of $I_\nu/N_\ion{H}{i}$ and $P_\nu/N_\ion{H}{i}$ for the 39\% of the sky having $N_\ion{H}{i} < 4.0\times 10^{20}\,{\rm cm^{-2}}$. The overall mean is plotted as a black dashed line while the black solid line is the running median. Right: Posterior distributions of fit parameters after applying the fitting formalism of Section~\ref{subsec:stat_model}. Best fit values are indicated with solid blue lines and are $m = -0.40_{-0.03}^{+0.03}$, $b = 4.51_{-0.01}^{+0.01}\times10^{22}$\,MJy\,sr$^{-1}$\,cm$^2$, $\sigma_I = 0.784_{-0.004}^{+0.004}\times10^{22}$\,MJy\,sr$^{-1}$\,cm$^2$.}
\label{fig:ip}
\end{figure*}

The most direct test of the sensitivity of the total dust intensity to the magnetic field orientation is the predicted anti-correlation between $I_\nu/N_\ion{H}{i}$ and $P_\nu/N_\ion{H}{i}$ (see Equation~\ref{eq:main}). In the left panel of Figure~\ref{fig:ip}, we present the 2D histogram of these quantities over all pixels having $N_\ion{H}{i} < 4\times10^{20}$\,cm$^{-2}$. The running median suggests a slight tendency for low values of $P_\nu/N_\ion{H}{i}$ to correspond to high values of $I_\nu/N_\ion{H}{i}$ and vice-versa, but the effect is not strong (Spearman rank coefficient $\rho = -0.15$).

We quantify the significance of the linear trend using Equation~\ref{eq:L_IP} and the formalism presented in Section~\ref{subsec:stat_model}. We indeed find a statistically significant preference for a negative slope, $m = -0.40\pm0.03$, even given the degeneracy with the fit intercept. The posterior distributions for each of the three fit parameters ($m$, $b$, and $\sigma_I$) are given in the right panel of Figure~\ref{fig:ip}.

Even if the negative correlation is robust, the slope is much shallower than the expected -1. However, positive correlations between $I_\nu/N_\ion{H}{i}$ and $P_\ion{H}{i}/N_\ion{H}{i}$ are easily induced from fluctuations in the dust to gas ratio, dust temperature, and dust opacity. Indeed, we find that $\sim10$\% variations in $T_d$ in Equations~\ref{eq:INH} and \ref{eq:PNH} are alone sufficient to reduce the slope to values comparable to what is observed. Further, systematic errors in separation of the Galactic dust from the CIB induces scatter in $I_\nu/N_\ion{H}{i}$ and potentially also correlations in this space. That the negative correlation induced by the orientation of ${\bf B}$ persists despite these complicating factors is indicative of the homogeneity of dust properties in the diffuse high latitude sky. A more robust correlation could likely be extracted with a higher fidelity map of $T_d$, and indeed the strength of this correlation may be a means of validating future dust temperature maps.

\subsection{$I_\nu/N_\ion{H}{i}$--$\mathcal{S}$ Correlation}
\label{subsec:I_S}

\begin{figure}
\centering
\includegraphics[width=0.48\textwidth]{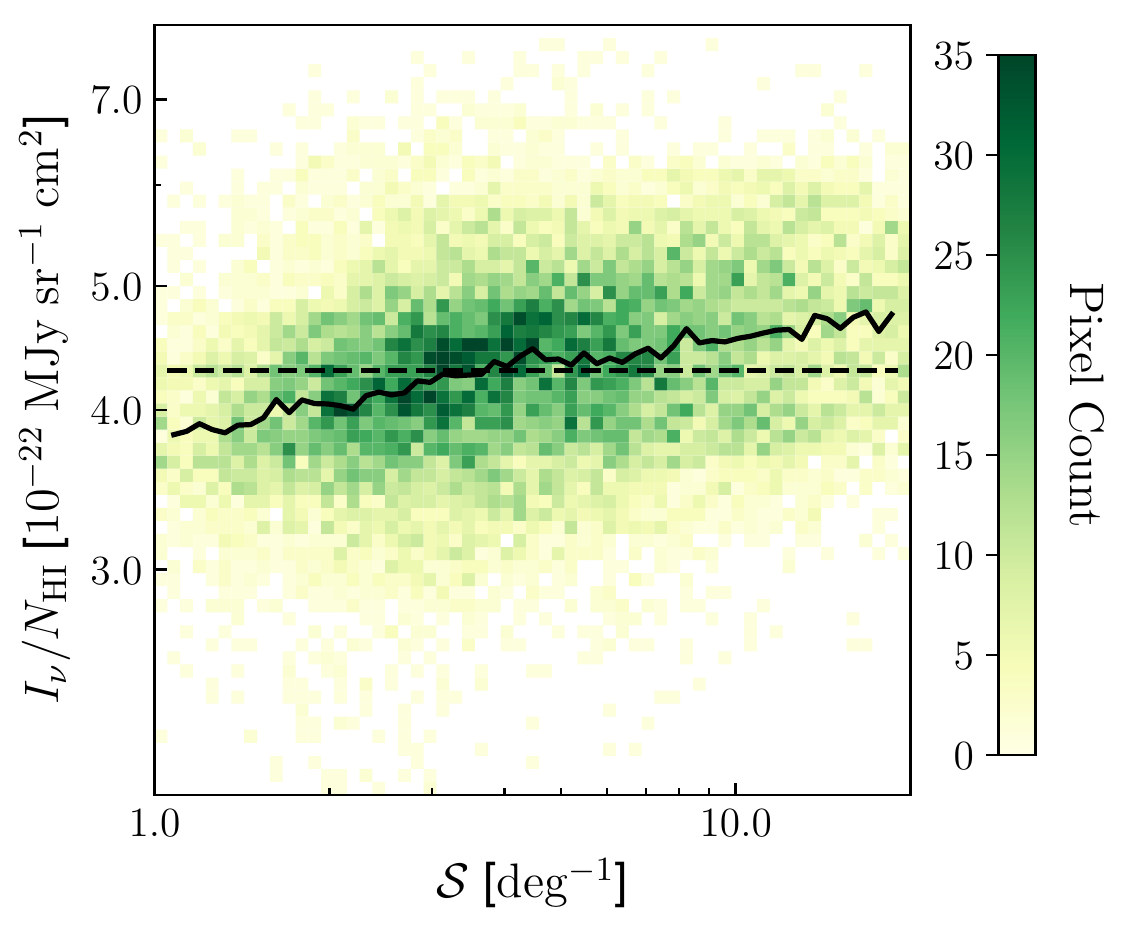}
\caption{2D histogram of $I_\nu/N_\ion{H}{i}$ and $\mathcal{S}$ for the 35\% of the sky having $N_\ion{H}{i} < 4.0\times 10^{20}\,{\rm cm^{-2}}$ and $1^\circ < \mathcal{S} < 20^\circ$. As in Figure~\ref{fig:ip}, the overall mean is plotted as a black dashed line while the black solid line is the running median. The Spearman rank coefficient $\rho = 0.29$.}
\label{fig:is}
\end{figure}

\begin{figure*}
\centering
\includegraphics[width=\textwidth]{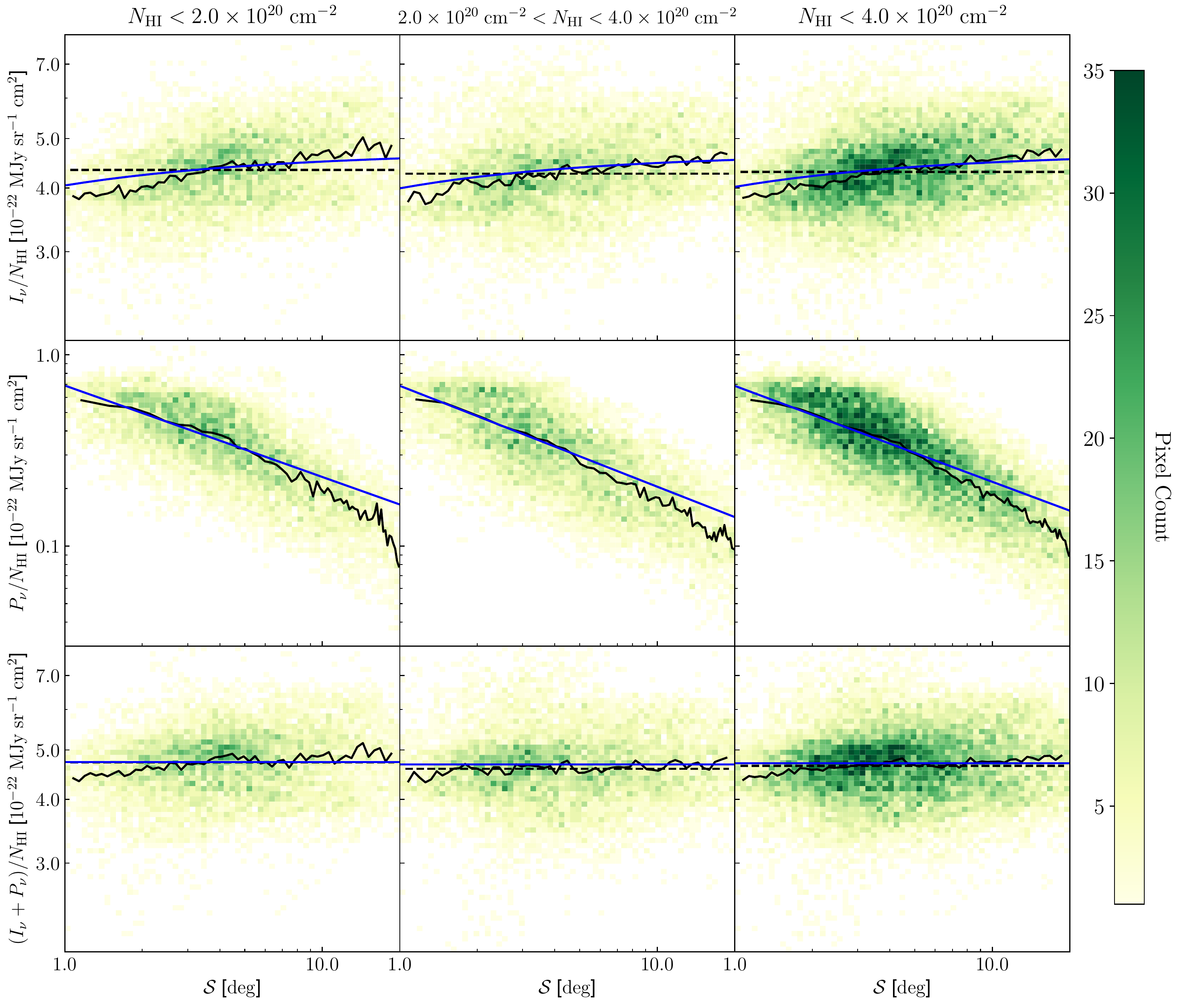}
\caption{2D histograms of $I_\nu/N_\ion{H}{i}$, $P_\nu/N_\ion{H}{i}$ and ${\mathcal{S}}$, with global means (black dashed) and running medians (black solid) plotted as in Figure~\ref{fig:ip}. The columns correspond to different $N_\ion{H}{i}$ thresholds: $N_\ion{H}{i} < 2.0\times10^{20}\,{\rm cm^{-2}}$ (left), $2.0\times10^{20}\,{\rm cm^{-2}} < N_\ion{H}{i} < 4.0\times10^{20}\,{\rm cm^{-2}}$ (middle), and $N_\ion{H}{i} < 4.0\times10^{20}\,{\rm cm^{-2}}$ (right). The top row illustrates the robustness of the $I_\nu/N_\ion{H}{i}$--$\mathcal{S}$ correlation to column density. The middle row presents the strong anti-correlation between $P_\nu/N_\ion{H}{i}$ and ${\mathcal{S}}$, similar to what has been observed using the polarization fraction \citep{Planck_2018_XII}. The bottom row demonstrates that summing $I_\nu/N_\ion{H}{i}$ and $P_\nu/N_\ion{H}{i}$ removes much of the correlation, as expected from Equation~\ref{eq:main}. The blue solid lines are the best joint fits to $I_\nu/N_\ion{H}{i}$, $P_\nu/N_\ion{H}{i}$, and $\mathcal{S}$ in each $N_\ion{H}{i}$ range. The best fit parameters are $A = 4.736\pm0.008$, $4.682\pm0.009$, and $(4.708\pm0.006)\times10^{22}$\,MJy\,sr$^{-1}$\,cm$^2$; $B = 0.691\pm0.005$, $0.689\pm0.005$, and $(0.688\pm0.003)\times10^{22}$\,MJy\,sr$^{-1}$\,cm$^2$; and $n = -0.478\pm0.005$, $-0.528\pm0.006$, and $-0.502\pm0.004$ in the left, middle, and right columns, respectively.}\label{fig:all2}
\end{figure*}

\begin{figure*}
\centering
\includegraphics[width=\textwidth]{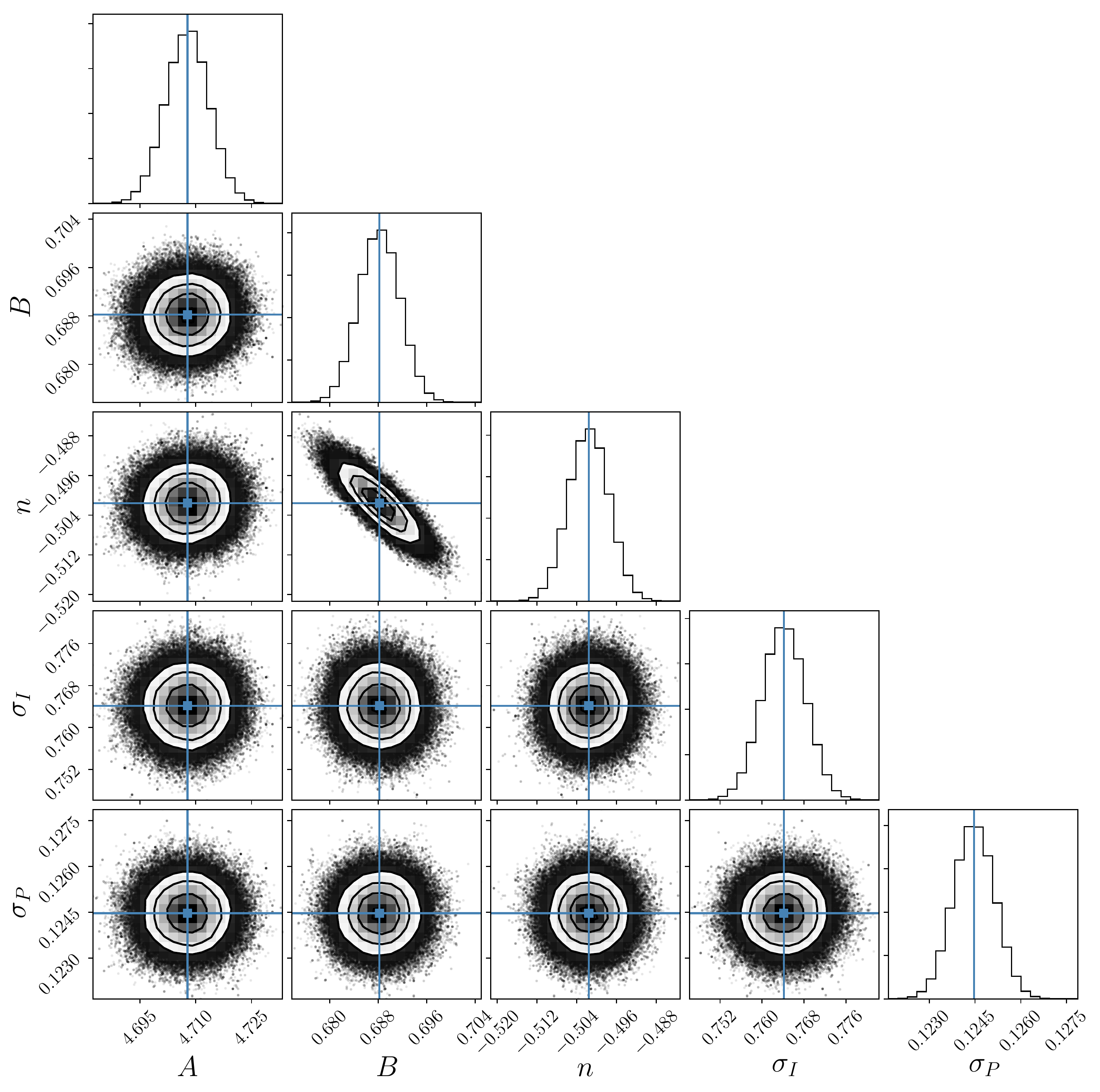}
\caption{Posterior distributions of fit parameters after applying the fitting formalism of Section~\ref{subsec:stat_model} to the $I_\nu/N_\ion{H}{i}$, $P_\nu/N_\ion{H}{i}$, and $S$ data. The fit was restricted to pixels having $N_\ion{H}{i} < 4\times10^{20}$\,cm$^{-2}$ and $1^\circ < \mathcal{S} < 20^\circ$. Best fit values are indicated with solid blue lines and are $A = \left(4.708\pm0.006\right)\times10^{22}$\,MJy\,sr$^{-1}$\,cm$^2$, $B = \left(0.688\pm0.003\right)\times10^{22}$\,MJy\,sr$^{-1}$\,cm$^2$, $n = -0.502\pm0.004$, $\sigma_I = \left(0.764\pm0.004\right)\times10^{22}$\,MJy\,sr$^{-1}$\,cm$^2$, and $\sigma_P = \left(0.124\pm0.001\right)\times10^{22}$\,MJy\,sr$^{-1}$\,cm$^2$.}\label{fig:all_triangle}
\end{figure*}

While the $I_\nu/N_\ion{H}{i}$--$P_\nu/N_\ion{H}{i}$ anti-correlation may be the most direct probe of the effect of the magnetic field orientation on the unpolarized dust emission, it is complicated by physical variations in dust properties that induce a positive correlation between these quantities, as discussed in the previous section. Therefore, we seek a quantity that is sensitive to the magnetic field orientation but insensitive to the dust temperature, dust to gas ratio, and other dust properties. In this section, we demonstrate that the polarization angle dispersion function $\mathcal{S}$ satisfies these criteria and robustly illustrates the sensitivity of $I_\nu/N_\ion{H}{i}$ to the angle $\psi$ between the Galactic magnetic field and the line of sight.

As discussed in Section~\ref{subsec:theory_S}, $\mathcal{S}$ is a measure of the variability of the polarization angle $\chi$ in a given region of the sky. Because $\mathcal{S}$ is computed only from polarization angles, it is not sensitive to the dust column density and has no {\it a priori} dependence on dust temperature. Further, changes in $\chi$ are more easily induced when ${\bf B}$ is oriented along the line of sight than when ${\bf B}$ is in the plane of the sky. Thus, high $\mathcal{S}$ is expected when ${\bf B}$ is along the line of sight and low $\mathcal{S}$ when in the plane of the sky. We therefore expect $I_\nu/N_\ion{H}{i}$ to be positively correlated with $\mathcal{S}$.

In Figure~\ref{fig:is}, we present the correlation between $I_\nu/N_\ion{H}{i}$ and $\mathcal{S}$ over the diffuse high latitude sky ($N_\ion{H}{i} < 4\times10^{20}$\,cm$^{-2}$). These quantities have a clear positive correlation (Spearman rank coefficient $\rho = 0.29$) despite considerable scatter. 

It is worth emphasizing the remarkable nature of this correlation. $I_\nu/N_\ion{H}{i}$ is the ratio of density tracers having no polarization information. $\mathcal{S}$ is computed purely from polarization angles, which contain no information about the density field. It is difficult to explain their evident correlation in any way other than through the orientation of the Galactic magnetic field.

We present a more detailed test of the correlation in Figure~\ref{fig:all2}. Each column in this figure corresponds to a different range of column densities, with the left column being $N_\ion{H}{i} < 2\times10^{20}$\,cm$^{-2}$, the middle $2\times10^{20}$\,cm$^{-2} < N_\ion{H}{i} < 4\times10^{20}$\,cm$^{-2}$, and the right $N_\ion{H}{i} < 4\times10^{20}$\,cm$^{-2}$. Thus, the first two columns are two independent sets of pixels on the sky. As in Figure~\ref{fig:is}, the top row presents the correlation of $\mathcal{S}$ with $I_\nu/N_\ion{H}{i}$. It is present at the same level in all three panels, demonstrating robustness to the range of $N_\ion{H}{i}$ considered.

The middle row of Figure~\ref{fig:all2} presents the correlation between $P_\nu/N_\ion{H}{i}$ and $\mathcal{S}$. \citet{Planck_2018_XII} demonstrated a clear anti-correlation between the polarization fraction $p$ and $\mathcal{S}$, and so the strong (Spearman rank coefficient $\rho \simeq -0.7$) anti-correlation seen here is expected.

The bottom row of Figure~\ref{fig:all2} correlates $(I_\nu+P_\nu)/N_\ion{H}{i}$ and $\mathcal{S}$. If the simple model described in Section~\ref{subsec:dust_emission} holds, then the positive correlation between $I_\nu/N_\ion{H}{i}$ and $\mathcal{S}$ should precisely balance the negative correlation between $P_\nu/N_\ion{H}{i}$ and $\mathcal{S}$. Indeed, the correlation with $\mathcal{S}$ is greatly reduced (Spearman rank coefficient $\rho \simeq 0.1$), though some residual correlation remains.

Figure~\ref{fig:all2} demonstrates that $\mathcal{S}$ is a proxy for the angle $\psi$ between ${\bf B}$ and the line of sight. If $\mathcal{S}$ were sensitive only to disorder in the magnetic field along the line of sight, then there would be no expected correlation with $I_\nu/N_\ion{H}{i}$. That this correlation exists and is of the same magnitude as the correlation with $P_\nu/N_\ion{H}{i}$ suggests that variation in $\psi$ is the {\it dominant} driver of the empirical anti-correlation between $\mathcal{S}$ and both $P_\nu/N_\ion{H}{i}$ (this work) and the polarization fraction $p$ \citep{Planck_2018_XII} across the high latitude sky.

Using the model presented in Equations~\ref{eq:I_S} and \ref{eq:P_S}, we can perform a joint fit to the $I_\nu/N_\ion{H}{i}$, $P_\nu/N_\ion{H}{i}$, and $\mathcal{S}$ data. We fit for the model parameters $A$, $B$, and $n$ as well as the intrinsic scatter $\sigma_I$ and $\sigma_P$ in $I_\nu/N_\ion{H}{i}$ and $P_\nu/N_\ion{H}{i}$, respectively, using the likelihood function in Equation~\ref{eq:L_IPS}. In addition to the column density threshold of $N_\ion{H}{i} < 4\times10^{20}$\,cm$^{-2}$, we restrict the fit to pixels having $1^\circ < \mathcal{S} < 20^\circ$ where the relationship between the polarization fraction and $\mathcal{S}$ is observed to be linear \citep{Planck_2018_XII}. The additional restriction on $\mathcal{S}$ reduces the usable sky fraction from 39\% to 35\%.

Over this sky area, corresponding to the right column of Figure~\ref{fig:all2}, we find $A = \left(4.708\pm0.006\right)\times10^{22}$\,MJy\,sr$^{-1}$\,cm$^2$, $B = \left(0.688\pm0.003\right)\times10^{22}$\,MJy\,sr$^{-1}$\,cm$^2$, $n = -0.502\pm0.004$, $\sigma_I = \left(0.764\pm0.004\right)\times10^{22}$\,MJy\,sr$^{-1}$\,cm$^2$, and $\sigma_P = \left(0.124\pm0.001\right)\times10^{22}$\,MJy\,sr$^{-1}$\,cm$^2$. The posterior distributions for each of the fit parameters are presented in Figure~\ref{fig:all_triangle}. The overall fit to the data is quite good, reinforcing the fact that the observed variation in $I_\nu/N_\ion{H}{i}$ is of the magnitude expected from orientation effects given the the observed variations in $P_\nu/N_\ion{H}{i}$. 

We find that $P_\nu/N_\ion{H}{i}$ scales approximately as $1/\sqrt{\mathcal{S}}$, somewhat shallower than the scaling of $p_\nu \propto S^{-0.67}$ found with BLASTPol observations in Vela~C \citep{Fissel+etal_2016} and significantly shallower than $p_\nu \propto S^{-1}$ found over the full sky with the same {\it Planck} 353\,GHz data as we employ \citep{Planck_2018_XII}. The $P_\nu/N_\ion{H}{i}$--$\mathcal{S}$ relation departs from a pure power law for $\mathcal{S} \gtrsim 10^\circ$. While the relation is expected to change character as $\mathcal{S}$ approaches the asymptotic value of $\pi/\sqrt{12} \simeq 52^\circ$ expected for uniform random variations in the polarization angles, that we observe this departure at smaller $\mathcal{S}$ than the full-sky {\it Planck} analysis may also point to spatial variability in the power law index $n$. As the relation is an empirical one depending on properties of MHD turbulence and the three-dimensional structure of the ISM \citep[see discussion in][]{Planck_2018_XII}, such variations are not unexpected. 

Indeed, for the other column density cuts presented in Figure~\ref{fig:all2}, we find slightly different values of our fit parameters. While $B$ is consistent with $0.689\times10^{22}$\,MJy\,sr$^{-1}$\,cm$^2$ in all $N_\ion{H}{i}$ ranges considered, $n$ varies from -0.478 for $N_\ion{H}{i} < 2\times10^{20}$\,cm$^{-2}$ to -0.528 for $N_\ion{H}{i}$ between 2 and $4\times10^{20}$\,cm$^{-2}$ with fitting uncertainty of about 0.005. Likewise, $A$ has best fit values of 4.736 and $4.682\times10^{22}$\,MJy\,sr$^{-1}$\,cm$^2$ in the two column density ranges, respectively, with an uncertainty of $0.009\times10^{22}$\,MJy\,sr$^{-1}$\,cm$^2$. 

As evident from the bottom row of Figure~\ref{fig:all2}, there remains residual correlation between $(I_\nu+P_\nu)/N_\ion{H}{i}$ and $\mathcal{S}$ that is not accounted for in the model. This could be attributed to a number of effects. First, the model considers only a single magnetic field orientation in a given pixel, both along the line of sight and within the beam. Variations in the field orientation affect the total and polarized intensities differently, complicating the simple linear correlation. Second, there can be true correlations between $\mathcal{S}$ and dust properties in our Galaxy. For instance, if $\mathcal{S}$ depends in part on the properties of MHD turbulence, and if those properties are found preferentially in regions of high or low dust temperature or dust to gas ratio, then such residual correlations could emerge.

\section{Discussion}
\label{sec:discussion}

We have demonstrated that the $\sim 10\%$ variation in the dust emission per grain induced by orientation effects is detectable in the 353\,GHz {\it Planck} data at a {\it statistical} level. Unlike the polarization angle which gives the projection of ${\bf B}$ onto the plane of the sky, this effect is sensitive only to the orientation of ${\bf B}$ relative to the line of sight, making it a complementary probe of the field orientation.

To progress from the statistical detection presented here to using this effect as a tracer of field orientation in specific sky regions, it will be necessary to improve the dust model fitting to extract reliable dust temperatures. Additional sensitive, high-frequency data, such as that provided by a mission like the Probe of Inflation and Cosmic Origins \citep[PICO,][]{Hanany+etal_2019}, would greatly improve existing SED fitting. In particular, high frequency polarization data is minimally contaminated by the CIB and thus may be more effective in constraining dust model parameters such as $T_d$ than the total intensity data.

In the nearer term, incorporation of \ion{H}{i} data into component separation, as has been demonstrated in CIB studies \citep{Planck_2013_XXX,Lenz+etal_2019} and is being explored in the context of CMB foregrounds \citep{Zhang+etal_2019}, has considerable potential for improving the fidelity of dust model fits. Further, such \ion{H}{i}-based model fits could explicitly account for the effect of viewing angle in the data model. 

Beyond improving determination of the dust temperature and other dust model parameters, \ion{H}{i} emission itself is a powerful tracer of the Galactic magnetic field. Linear filamentary structures seen in \ion{H}{i} emission have been shown to align strongly with the local magnetic field \citep{Clark+etal_2015}, allowing the magnetic field orientation to be traced both as a function of position on the sky as well as \ion{H}{i} velocity (Clark \& Hensley, in prep.). A synthesis of far-infrared and \ion{H}{i} emission could therefore in principle enable mapping of the full three-dimensional orientation of the Galactic magnetic field over the sky. With existing stellar distances from {\it Gaia} \citep{Gaia_2016} and upcoming starlight polarization measurements from PASIPHAE \citep{Tassis+etal_2018}, such Galactic magnetic field models could be translated from magnetic field orientations in \ion{H}{i} velocity-space to the full distribution of dust and magnetic field orientations in the three spatial dimensions.

The line of sight component of the Galactic magnetic field can be constrained in ways other than the one presented here. For instance, the dust polarization fraction $p_\nu$ is sensitive to the inclination of ${\bf B}$ (see Equation~\ref{eq:pfrac}), but is complicated by non-uniformity in the magnetic field orientation along the line of sight \citep{Clark_2018}. Joint modeling of $p_\nu$ and $I_\nu/N_\ion{H}{i}$ can help overcome the different limitations of each diagnostic and therefore also quantitatively distinguish between the effects of magnetic field orientation and beam/line-of-sight depolarization. Additionally, Faraday rotation and Zeeman splitting have both been used to measure the line of sight component of ${\bf B}$, though they may preferentially probe different ISM phases than dust-based tracers.

Dust emission in total intensity and polarization are tightly coupled in ways straightforwardly captured in parametric models, as described in Section~\ref{subsec:dust_emission}. As the quest for the detection of primordial B-modes in the polarized CMB pushes to ever-increasing precision, the importance of accurate modeling and subtraction of dust foregrounds is paramount \citep[see discussion in][]{BICEP2_2018}. Effects like the one discussed in this work enable other datasets, such as dust intensity and \ion{H}{i} emission, to be brought to bear on modeling dust polarization. Even if not accounted for explicitly, this effect can validate dust temperatures and dust column densities that emerge from model fitting. As we are pushed to higher fidelity modeling of dust foregrounds, we should at the same time be learning about the structure of the magnetized ISM in increasing detail.

\section{Conclusions}
\label{sec:conclusions}

The principal conclusions of this work are as follows:

\begin{enumerate}
    \item Because dust grains are aspherical, the effective cross section and thus total emission from dust depends on the viewing angle. We demonstrate that this induces a direct anti-correlation between the total dust intensity per $N_\ion{H}{i}$ and the polarized dust intensity per $N_\ion{H}{i}$ (Equation~\ref{eq:main}).
    \item In the low column density sky ($N_\ion{H}{i} < 4\times10^{20}$\,cm$^{-2}$), we find that application the GNILC dust temperature map does not improve the correlation between the 353\,GHz dust intensity and the \ion{H}{i} column density as expected. We suggest this arises from parameter degeneracies in the low signal to noise regime. The tests outlined in Section~\ref{subsec:Tdust} can provide a means of validating future dust temperature maps.
    \item We find a robust positive correlation between $I_\nu/N_\ion{H}{i}$ and the polarization angle dispersion function $\mathcal{S}$ as predicted from the orientation effect. This suggests that observed variations in $\mathcal{S}$ are driven largely by the variations in the angle between the Galactic magnetic field and the line of sight.
    \item We find that the positive correlation between $I_\nu/N_\ion{H}{i}$ and $\mathcal{S}$ is largely compensated by the anti-correlation between $P_\nu/N_\ion{H}{i}$ and $\mathcal{S}$, as predicted by Equation~\ref{eq:main}.
    \item We argue that the variation in dust emission per $N_\ion{H}{i}$ can be used to probe the full 3D magnetic field orientation vector, complementing the plane of sky orientation provided by the dust polarization angle.
    \item We suggest this effect can also be exploited in component separation in CMB experiments, particularly since the formalism developed in Section~\ref{subsec:dust_emission} provides a physically-motivated means of connecting observations of total intensity and polarization.
\end{enumerate}

\acknowledgments
{It is a pleasure to thank Fran\c{c}ois Boulanger, Susan Clark, Bruce Draine, Laura Fissel, and Vincent Guillet for stimulating discussions and helpful suggestions. This research was carried out in part at the Jet Propulsion Laboratory, California Institute of Technology, under a contract with the National Aeronautics and Space Administration.}

\software{Astropy \citep{Astropy,Astropy_2}, {\it corner} \citep{corner}, \texttt{emcee} \citep{ForemanMackey+etal_2013}, HEALPix \citep{Gorski+etal_2005}, \texttt{healpy} \citep{healpy}, Matplotlib \citep{Matplotlib}, NumPy \citep{NumPy}}

\facility{Effelsberg, Parkes, Planck}

\bibliography{dust} 
\end{document}